\documentclass[fleqn,twoside,twocolumn,nofootinbib]{revtex4} 
\usepackage[sec]{ujp} 
\begin{document}
\title[NONLINEAR PLASMA DIPOLE OSCILLATIONS]
{NONLINEAR PLASMA DIPOLE OSCILLATIONS\\ IN SPHEROIDAL METAL NANOPARTICLES}%
\author{P.M.~Tomchuk}
\affiliation{Institute of Physics, Nat. Acad. of Sci. of Ukraine}
\address{46, Prosp. Nauky, Kyiv 03680, Ukraine}
\email{ptomchuk@iop.kiev.ua}
\author{D.V.~Butenko}
\affiliation{Institute of Physics, Nat. Acad. of Sci. of Ukraine}
\address{46, Prosp. Nauky, Kyiv 03680, Ukraine}
\email{ptomchuk@iop.kiev.ua} \udk{???} \pacs{41.20.Cv, 73.20.Mf,\\[-3pt]
78.67.-n} \razd{\secix}

\setcounter{page}{1110}%
\maketitle


\begin{abstract}
The theory of nonlinear dipole plasma oscillations generated in a
metal spheroidal nanoparticle by a laser-wave field has been
developed. Approximate (to within the cubic term) analytic
expressions for the nanoparticle dipole moment have been obtained in
the case where the laser field is oriented along the spheroid rotation
axis.
\end{abstract}

\section{Introduction}

When the center of masses of the electron subsystem in a metal nanoparticle is
shifted with respect to the center of masses of the ion subsystem,
there emerges an electrostatic force, which counteracts their spatial
separation. This force may invoke dipole plasma oscillations in metal
nanoparticles. In the first approximation, it is proportional to the relative
displacement of the electron and ion centers of masses. If the displacement grows
further, the electrostatic force starts to depend nonlinearly on this
separation, which results in the appearance of nonlinear dipole plasma
oscillations. These nonlinear plasma oscillations in metal nanoparticles were
studied in works \cite{1,2,3}. In works \cite{1,2}, plasma oscillations were
considered in the continual approximation, and the microscopic approach was
taken as a basis in work \cite{3}. In all cited works, the shape of a metal
nanoparticle was assumed spherical.

Our work is devoted to the development of the theory of nonlinear plasma
oscillations in metal nanoparticles of ellipsoidal shape. It should be
emphasized that the results of the theory of plasma resonances in asymmetric
metal nanoparticles cannot be reduced to small corrections to the results
known for spherical particles, but have fundamental differences. In
particular, already in the linear approximation, a spherically symmetric metal
particle has one plasma resonance, whereas a spheroidal particle has two plasma resonances and
an ellipsoidal particle has three ones. Therefore, the task aimed at
developing the nonlinear theory of dipole plasma oscillations in asymmetric
metal nanoparticles remains challenging and interesting for today.

\section{Formulation of the Problem}

We consider the problem of oscillations in metal
nanoparticles in the continual approximation and take, as a basis,
the hydrodynamic equations for the
electron density $n_{e}(\mathbf{r},t)$ and the electron velocity
$\boldsymbol{\upsilon}(\mathbf{r},t)$ similarly to work \cite{2}:
\begin{equation}
\frac{\partial n_e }{\partial t}+\boldsymbol{\nabla
}\;(n_e \boldsymbol{\upsilon})=0,\label{eq1}%
\end{equation}
\[
 \frac{\partial \boldsymbol{\upsilon}}{\partial
t}+(\boldsymbol{\upsilon}\;\boldsymbol{\nabla })\boldsymbol{\upsilon
}=\frac{\mathbf{F}}{m_e }\equiv \frac{1}{m_e }\times
\]
\begin{equation}\label{eq2}
\times\left\{ {-e\,\mathbf{E}_L +e\left[ {\boldsymbol{\nabla }\Phi
_e +\boldsymbol{\nabla }\,\Phi _i }
\right]\,-\frac{\boldsymbol{\nabla }\,p}{n_e }} \right\}.
\end{equation}
In Eq.~(\ref{eq2}), $\mathbf{F}(\mathbf{r},t)$ is the total force
that acts on the electron liquid. It is composed of the action of
the electric field generated by the laser wave, $\mathbf{E}_{L}$,
and the action of the gradients of electron, $\Phi_{e}$, and ion,
$\Phi_{i}$, potentials, and the pressure, $p$, gradient. In the
dipole approximation, the field $\mathbf{E}_{L}$ is considered
spatially uniform within a nanoparticle.

Let us introduce a vector that characterizes the position of the center of
masses of the electron subsystem,
\begin{equation}
\mathbf{u}\,(t)=N_{e}^{-1}\int{d^{3}\,r\,n_{e}(\mathbf{r},t)\,\mathbf{r}%
},\label{eq3}%
\end{equation}
where $N_{e}$ is the total number of electrons in the metal nanoparticle.
With regard for Eq.~(\ref{eq1}), the equation of motion for the center of
masses of the electron subsystem looks like
\[
N_{e}\;\frac{d^{2}\mathbf{u}}{dt^{2}}=\int{d^{3}r\,\frac{\partial^{2}n_{e}%
\,(\mathbf{r},t)}{\partial
t^{2}}}\;\mathbf{r}=
\]
\begin{equation}
=\int{d^{3}r\;n_{e}(\mathbf {r},t)\;\left(
{\frac{\partial\boldsymbol{\upsilon}}{\partial
t}+(\boldsymbol{\upsilon}\,\boldsymbol{\nabla
})\,\boldsymbol{\upsilon}}\right)  }.\label{eq4}%
\end{equation}
Rewriting it in the form
\begin{equation}
m_{e}\;\mathbf{u}=N_{e}^{-1}\langle\,\mathbf{F}\,\rangle,\label{eq5}%
\end{equation}
and using Eq.~(\ref{eq2}), we obtain
\[
\langle \mathbf{F}\rangle =-e\;E_L (t)N_e +
\]
\begin{equation}
\label{eq6}
 +\!\!\int\!\! {d^3r\,\left\{ {e\,\left[ {\boldsymbol{\nabla}\Phi _e (\mathbf{r},t)+
 \boldsymbol{\nabla}\Phi _i (\mathbf{r},t)} \right]\,n_e (\mathbf{r},t)-\boldsymbol{\nabla}p\,(\mathbf{r},t)} \right\}}.
\end{equation}
Above, we have briefly reproduced the approach to plasma dipole oscillations
in metal nanoparticles used in work \cite{2}. The authors of work \cite{2},
by making the required estimations, adopted an approximation, whose
essence is the assumption that the electron subsystem of a nanoparticle shifts
as a whole (without deformations) together with the center of masses of electrons.
This allows us to put
\[
n_{e}\,(\mathbf{r},t)=n_{e}^{(0)}\,(\,|\mathbf{r}-\mathbf{u}\,(t)|),
\]%
\begin{equation}
\Phi_{e}\,(\mathbf{r},t)=\Phi_{e}^{(0)}\,(\,|\mathbf{r}-\mathbf{u}\,(t)|),\label{eq7}%
\end{equation}%
\[
\mathbf{p}\,(\mathbf{r},t)=p_{0}\,(\,|\mathbf{r}-\mathbf{u}\,(t)|).
\]
We note that, at the
thermodynamic equilibrium (i.e., in the absence of
$\mathbf{E}_{L}\,(t)$), the equality
\begin{equation}
e\,\left[ {\boldsymbol{\nabla }\Phi_{e}^{(0)}+\boldsymbol{\nabla
}\Phi_{i}^{(0)}}\right]
-\frac{\boldsymbol{\nabla }\,\rho^{(0)}}{n_{e}}=0\label{eq8}%
\end{equation}
is valid. In this case, Eqs.~(\ref{eq6}) and (\ref{eq7}) yield
\[
 \langle \mathbf{F}\,\rangle =-e\,\mathbf{E}_L \,(t)\,N_e +\int {d^3r\,e\,
 \boldsymbol{\nabla }\,\Phi _i^{(0)} (\mathbf{r})} n_e^{(0)} (\vert
 \mathbf{r}-\mathbf{u}(t)\vert )=
\]
\begin{equation}
\label{eq9}
 =-e\,\mathbf{E}_L (t)\;N_e +\frac{\partial }{\partial \mathbf{u}}\,\int
{d^3\,r\,e\,\Phi _i^{(0)} (\mathbf{r})\;n_e^{(0)} (\vert \mathbf{r}
-\mathbf{u}\vert )}.
\end{equation}
We use this formula as the basic one.

In contrast to the previous works \cite{1,2,3} where plasma nonlinear
oscillations in spherically symmetric metal
nanoparticles were considered, we analyze asymmetric nanoparticles. Let a metal
nanoparticle have the shape of an ellipsoid of revolution (spheroid).
Let the coordinate $0Z$ axis be oriented along the spheroid symmetry
axis. In addition, we suppose that the laser field $\mathbf{E}_{L}(t)$,
as well as a shift of the center of masses of the electron subsystem
induced by the field, is directed along the $0Z$ axis. To emphasize
this fact, we use the notation
\begin{equation}
\mathbf{u}\equiv\mathbf{z}_{0}\label{eq10}%
\end{equation}
in what follows. It is worth noting that, if the particle's shape differs from the
sphere, and the laser field that generates plasma oscillations is not oriented
strictly along the symmetry axis, the various plasma resonances are coupled with
one another in the nonlinear approximation, and the problem becomes
incredibly complicated. Our calculations given below show that even the
simplest model presented above and allowing for deviations from spherical
symmetry is already capable to produce qualitatively new results, as
compared with the spherical case.

Up to now, except for formulas (\ref{eq7}) and (\ref{eq9}), the nanoparticle
symmetry has not been specified. Similarly to work \cite{2}, we
adopt that the electron subsystem shifts as a whole (without deformations)
together with its center of masses. Hence, we adopt that
\begin{equation}
n_{e}(\mathbf{r},t)=n_{0}\,\Delta_{0}\,(R(\theta)-|\,\mathbf{r}-\mathbf{z}%
_{0}|),\label{eq11}%
\end{equation}
where $n_{0}$ is the concentration of electrons, and $\Delta_{0}(x)$ is the
step-like function
\begin{equation}
\label{eq12} \Delta _0 (x)={\left\{ {\begin{array}{l}
 1,\quad x>0, \\
 0,\quad x<0 .\\
 \end{array}} \right.}
\end{equation}
The function $R(\theta)$ defines the
spheroid surface (to be more specific, let the spheroid be prolate),
\begin{equation}
R\,(\theta)=\frac{R_{\bot}}{\sqrt{1-e_{p}^{2}\,\cos^{2}\theta}},\label{eq13}%
\end{equation}
where $e_{p}$ is the spheroid eccentricity,
\begin{equation}
e_{p}^{2}=\;\left\vert
{\,\frac{R_{\perp}^{2}}{R_{\parallel}^{2}}\;-1\;}\right\vert
,\label{eq14}%
\end{equation}
$\theta$ is the angle between the axis $0Z$ and the
$\mathbf{R}(\theta)$ on the spheroid surface, and $R_{||}$ and
$R_{\bot}$ are the longitudinal (along the symmetry axis) and
transverse, respectively, curvature radii.

Supposing, in analogy with Eq.~(\ref{eq7}), that the structures of
functions $\Phi_{e}(\mathbf{r},t)$ and $p\,(\mathbf{r},t)$ are
similar to that of the function $n_{e}(\mathbf{r},t)$ given by formula
(\ref{eq11}), we obtain
\[
\langle \mathbf{F}\rangle =-e\,\mathbf{E}_L (\,t\,)\;N_e +n_0 \times
\]
\begin{equation}
\label{eq15} \times\int {d^3r\;e\;\bar{\nabla}\;\Phi _i^{(0)}
(\,\mathbf{r}\,)\;\Delta _0 \;(R(\theta )-\vert
\,\mathbf{r}-\mathbf{z}_0 \,\vert )}
\end{equation}
instead of formula (\ref{eq9}). Hence, in order to determine the force $\left\langle
\mathbf{F}\right\rangle $ that counteracts a displacement of the
electron subsystem in a spheroidal metal nanoparticle along the
axis $0Z$, we have to determine the ionic electrostatic potential
$\Phi_{i}^{(0)}$. This will be done in the next section.

\section{Electrostatic Potential of a Charged Spheroid}

The electrostatic potential generated by the ion core in a spheroidal
nanoparticle looks like
\begin{equation}
\Phi_{i}^{(0)}=\int\limits_{V}{\frac{\rho_{i}\,d^{3}r}{|\;\mathbf{r}-{\mathbf{r}%
}^{\prime}\;|}}.\label{eq16}%
\end{equation}
Let the density of ion charges be uniformly distributed over the volume $V$,
\begin{equation}
\label{eq17} \rho _i ={e\,N_i \,Z_{i\;} }/ V={\mathrm{const}},
\end{equation}
where $N_{i}$ is the number of ions, and $Z_{i}$ is the charge multiplicity. Taking the spheroid
symmetry and the charge uniformity
into account, Eq.~(\ref{eq16}) takes the form
\begin{equation}
\Phi_{i}^{(0)}=\rho_{i}\int\limits_{0}^{2\pi}{d\,{\varphi}^{\prime}%
\int\limits_{0}^{\pi}{d\,{\theta}^{\prime}\,\sin\,{\theta}^{\prime}}}%
\int\limits_{0}^{R\,({\theta}^{\prime})}{\frac{d{r}^{\prime}\,{r}^{\prime
}{}^{2}}{|\mathbf{r}-{\mathbf{r}}^{\prime}|}}.\label{eq18}%
\end{equation}
To carry out the integration in Eq.~(\ref{eq18}), it is expedient to
make the expansion
\begin{equation}
\frac{1}{|\,\mathbf{r}-{\mathbf{r}}^{\prime}\,|}=\sum\limits_{n=0}^{\infty
}{{P}_{n}\,(\cos\nu)}\left\{  {%
\begin{array}
[c]{l}%
\frac{1}{r}\left(  {\frac{{r}^{\prime}}{r}}\right)  ^{n}\quad \mathrm{at}\quad{r}%
^{\prime}<r,\\
\frac{1}{{r}^{\prime}}\left(  {\frac{r}{{r}^{\prime}}}\right)  ^{n}\quad
\mathrm{at}\quad{r}^{\prime}>r,\\
\end{array}
}\right. \label{eq19}%
\end{equation}
where $\nu$ is the angle between the vectors $\,\mathbf{r}$ and ${\mathbf{r}%
}^{\prime}$, and apply the relation \cite{4}
 \[
  {P}_n \,(\cos \,\nu )={P}_n \,(\cos {\theta }')\;{P}_{n\,} (\cos \theta )+
\]
\begin{equation}
\label{eq20} +2\sum\limits_m {\frac{(n-m)!}{(n+m)!}\cos ({\varphi
}'-\varphi )\,{P}_n^m \,(\cos {\theta }')}\, {P}_n^m (\cos \theta ).
\end{equation}
The angles $({\varphi}^{\prime},{\theta}^{\prime})$ and
$(\varphi,\theta)$ in Eq.~(\ref{eq20}) describe the spatial
orientations of the vectors $\mathbf
{r}^{\prime}$ and ${\mathbf{r}}$, respectively. When substituting Eq.~(\ref{eq20}%
) in Eq.~(\ref{eq18}) and integrating the result obtained over ${\varphi
}^{\prime}$, the second term in Eq.~(\ref{eq20}) is nulled.

According to Eq.~(\ref{eq13}), the quantity $R({\theta}^{\prime})$ in
Eq.~(\ref{eq18}) satisfies the condition
\begin{equation}
R_{\perp}\leq R\,({\theta}^{\prime})\leq R_{\parallel}. \label{eq21}%
\end{equation}
Therefore, it is expedient to consider the integral over
$\mathbf{r}^{\prime}$ in three cases:
\[
 a)~r\,\leq\,R_{\perp},
 \]
 \[
 b)~R_{\perp}\leq
r\,\leq\,R_{\parallel},~~ \mathrm{and}
\]
\begin{equation}
c)~R_{\parallel}\,\,\leq\,\,r.
\end{equation}
At $r\,\leq\,R_{\perp}$, in accordance with
Eqs.~(\ref{eq18})--(\ref{eq20}), we obtain
\[
\Phi_{i}^{(0)}=2\pi\,\rho_{i}\sum\limits_{n=0}^{\infty}{P_{n}%
\,(\cos\theta)\int\limits_{0}^{\pi}{d{\theta}^{\prime}\,\sin{\theta}%
^{\prime}{P}_{n}\,(\cos{\theta}^{\prime})}}\times
\]
\begin{equation}%
\times\left\{  {\frac{1}{r}\int\limits_{0}^{r}{d{r}^{\prime}\;{r}^{\prime}%
{}^{2}\left(  {\frac{{r}^{\prime}}{r}}\right)  ^{n}+\int\limits_{r}%
^{R({\theta}^{\prime})}{d{r}^{\prime}\;{r}^{\prime}\;\left(  {\frac{r}%
{{r}^{\prime}}}\right)  ^{n}}}}\right\}.
\label{eq22}%
\end{equation}
The further calculation of the integrals in formula (\ref{eq22}) has no
difficulties. A more difficult situation arises in the case $R_{\bot}\leq
r\leq R_{\Vert}$, because, in accordance with Eq.~(\ref{eq21}), the variable
$r^{\prime}$ changes in the same interval. Therefore, depending on the angle
$\theta^{\prime}$, the maximum value of $r^{\prime}$ can be both larger and
smaller than $r$ (see Fig.~1). It is expedient to introduce an angle
$\theta_{1}$, at which the ellipsoid and the sphere of radius $r$ intersect;
in other words,
\begin{equation}
R\,({\theta}^{\prime})=R\,(\theta_{1})=r.\label{eq23}%
\end{equation}
In view of Eq. (\ref{eq13}), Eq. (\ref{eq23}) yields
\begin{equation}
\cos\theta_{1}=\frac{1}{e_{\rho}}\left\{  {1-\left(  {\frac{R_{\bot}}{r}^{2}%
}\right)  }\right\} ^{1/ 2}.%
\label{eq24}%
\end{equation}

Now, let us decompose the integral over $\theta^{\prime}$ in
Eq.~(\ref{eq18}) as follows:
\begin{equation}
\int\limits_{0}^{\pi}{d{\theta}^{\prime}}\ldots=\int\limits_{0}^{\theta_{1}%
}{d{\theta}^{\prime}\ldots+\int\limits_{\theta_{1}}^{\pi-\theta_{1}}{d{\theta
}^{\prime}}}\ldots+\int\limits_{\pi-{\theta}_{1}}^{\pi}{d{\theta
}^{\prime}\ldots}~.\label{eq25}%
\end{equation}
Figure~1 demonstrates that $r^{\prime}$ can be both larger and smaller
than $r$ in the intervals $0<\theta^{\prime}<\theta_{1}$ and $\pi-\theta
_{1}<\theta^{\prime}<\pi$. At the same time, $r^{\prime}$ is always smaller
than $r$ in the interval $\theta_{1}<\theta^{\prime}<\pi-\theta_{1}$.
Substituting expansion (\ref{eq19}) in Eq.~(\ref{eq18}) and dividing the
integration interval in accordance with procedure (\ref{eq25}), we obtain
\[
 \Phi _i^{(0)} =2\pi \rho _i \sum\limits_{n=0}^\infty {P_n (\cos
\theta )}\Biggl\{ \int\limits_0^{\theta _1 } {d{\theta }'\sin
{\theta }' {P}_n (\cos {\theta }')} \times
\]
\[
 \times \Bigg[ \frac{1}{r}\int\limits_0^r {d{r}'{r}'^2\left(
{\frac{{r}'}{r}} \right)^n}+\int\limits_r^{R({\theta }')}
{d{r}'{r}'\left( {\frac{r}{{r}'}} \right)^n} \Bigg]+
\]
\[
 +\int\limits_{\theta _1 }^{\pi -\theta _1 } {d{\theta }'\sin {\theta
}'P_n (\cos {\theta }')\frac{1}{r}}\int\limits_0^{R({\theta }')}
{d{r}'{r}'^2\left( {\frac{{r}'}{r}} \right)^n+}
\]
\[
 +\int\limits_{\pi -{\theta }_1 }^\pi {d{\theta }'\sin {\theta
}'{P}_n (\cos {\theta }')}\times
\]
\begin{equation}
\label{eq26} \times\Bigg[ \frac{1}{r}\,\int\limits_0^r
{d{r}'{r}'^2\left( {\frac{{r}'}{r}}
\right)^n}+\int\limits_r^{R({\theta }')} {d{r}'{r}'\left(
{\frac{r}{{r}'}} \right)^n }  \Bigg] \Biggr\}.
 \end{equation}
Making substitutions of the type
${\theta}^{\prime}\div\pi-{\theta }^{\prime}$ in the last term of
Eq.~(\ref{eq26}), the whole expression (\ref{eq26}) can be expressed
in the form
\[
\Phi_{i}^{(0)}=2\pi \rho_{i}\sum\limits_{n=0}^{\infty}{P_{n}
\,(\cos\theta)}\Bigg\{ \int\limits_{0}^{\theta_{1}}{d{\theta
}^{\prime}\,\sin{\theta}^{\prime}}\times
\]
\[
\times\left[{P}%
_{n}(\cos{\theta}^{\prime})+{P}_{n}(-\cos{\theta}^{\prime })\right]
\times
\]
\[
\times\Bigg[  \frac{1}{r}\int\limits_{0}^{r}{d{r}^{\prime}\,{r}^{\prime}%
{}^{2}\left(  {\frac{{r}^{\prime}}{r}}\right)  ^{n}}+\int\limits_{r}%
^{R({\theta}^{\prime})}{d{r}^{\prime}\,{r}^{\prime}\left(  {\frac{r}%
{{r}^{\prime}}}\right)  ^{n}}\Bigg]  +
\]
\[
+\int\limits_{\theta_{1}}^{\pi/2_1}{d{\theta}^{\prime}\sin{\theta}^{\prime}}\left[
P_{n}(\cos{\theta}^{\prime})+{P}_{n}(-\cos{\theta}^{\prime} )\right]
\times
\]
\begin{equation}%
\times
\frac{1}{r}\int\limits_{0}^{R({\theta}^{\prime})}{d{r}^{\prime}
{r}^{{\prime}^{2}}\left(  \frac{{r}^{\prime}}{r}\right)^{n}}\Bigg\}.
\label{eq27}%
\end{equation}
One can see that all terms with odd powers of $n$ disappear from sum
(\ref{eq27}). Expression (\ref{eq27}) describes $\Phi_{i}^{(0)}$ in
the range $R_{\perp}\leq r\leq R_{\parallel}$.

\begin{figure}
\includegraphics[width=5cm]{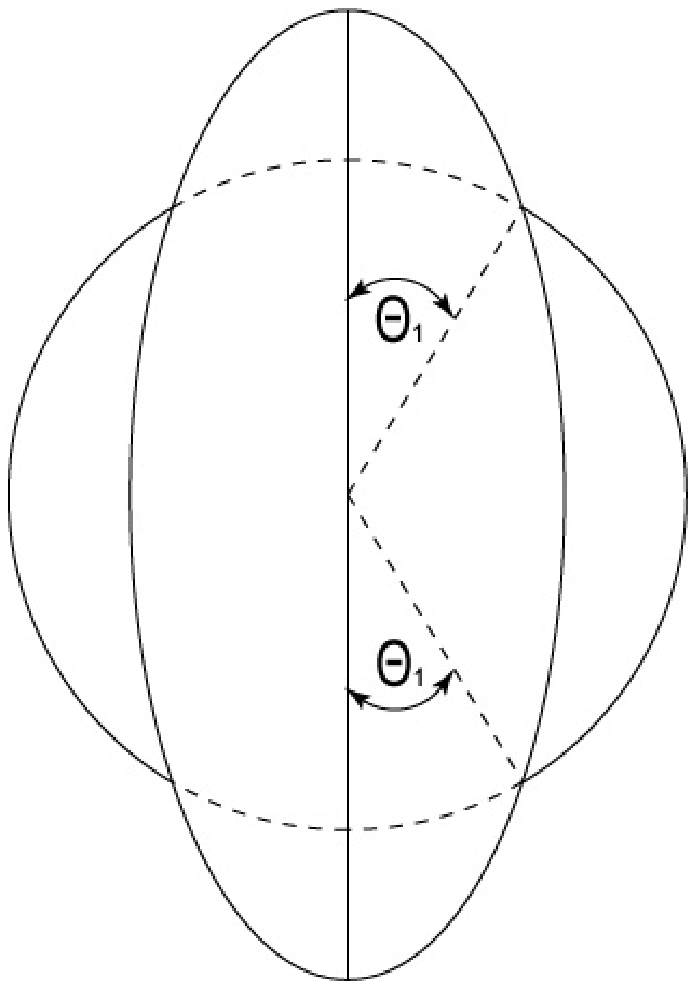}
\vskip-3mm\caption{  }
\end{figure}

At last, let us consider the case $r\geq R_{\parallel}$, where
$r^{\prime }<r$. From Eqs.~(\ref{eq18}) and (\ref{eq19}), we
obtain
\[
 \Phi _i^{(0)} =2\pi ,\rho _i \sum\limits_{n=0}^\infty P_n \,(\cos \theta )\int\limits_0^\pi
{d{\theta }'\,\sin {\theta }'\times{P}_n (\cos {\theta }')}\times
\]
\begin{equation}
\label{eq28}  \times \frac{1}{r}\int\limits_0^{R({\theta }')}
{d{r}'\,{r}'^{2\;}\left( {\frac{{r}'}{r}} \right)} ^n.
\end{equation}
As is seen from expressions (\ref{eq22}), (\ref{eq27}), and
(\ref{eq28}),
we can write
\[
\Phi_{i}^{(0)}=\sum\limits_{n=0}^{\infty}{{P}_{n}(\cos\theta
)\Psi(n)=\Psi(0)+}
\]
\begin{equation}
{+{P}_{2}(\cos\theta)\Psi(\ref{eq2})+{P}_{4}%
(\cos\theta)\Psi(\ref{eq4})+\cdots} \label{eq29}%
\end{equation}
for the whole range of variation of the vector $\mathbf{r}$. Integrating in Eqs.~(\ref{eq22}), (\ref{eq27}), and (\ref{eq28}), we
obtain the expressions for the coefficients $\Psi(n)$. In particular,
at $r\leq R_{\perp}$, we find
\begin{equation}
\Psi(0)=2\pi\rho_{i}\left\{  {-\frac{r^{2}}{3}+\frac{R_{\perp}^{2}%
}{2e_{p}}\ln\left(  {\frac{1+e_{p}}{1-e_{p}}}\right)  }\right\} ;~~
r\leq
R_{\bot}, \label{eq30}%
\end{equation}
using Eq.~(\ref{eq22}). From Eq.~(\ref{eq27}), we obtain that, at $R_{\bot}\leq r\leq
R_{\parallel}$,
\[
\Psi (0)=2\,\pi \rho _i \biggl\{ \frac{r^2}{3}(\cos \theta _1 -1)-
\]
\[
-\frac{R_\perp ^2 }{2e_p }\ln \,\left( {\frac{1+e_p \cos \theta _1
}{1-e_p \cos \theta _1 }\;\frac{1-e_p }{1+e_p }}
\right)+\frac{2R_\perp ^2 }{3}\cos \theta _1  \biggr\},
\]
\begin{equation}
R_{\perp}\leq r\leq R_{\parallel}. \label{eq31}%
\end{equation}
At last, Eq.~(\ref{eq28}) implies that, at $r\geq
R_{\parallel}$,
\begin{equation}
\Psi(0)=\frac{4}{3}\pi\rho_{i}R_{\perp}^{2}{R_{\parallel}}%
R_{\parallel}/r,\quad r\ge R_{\parallel }
\label{eq32}%
\end{equation}
Similar expressions for $\Psi(2)$ are given in Appendix.

\begin{figure}
\includegraphics[width=5cm]{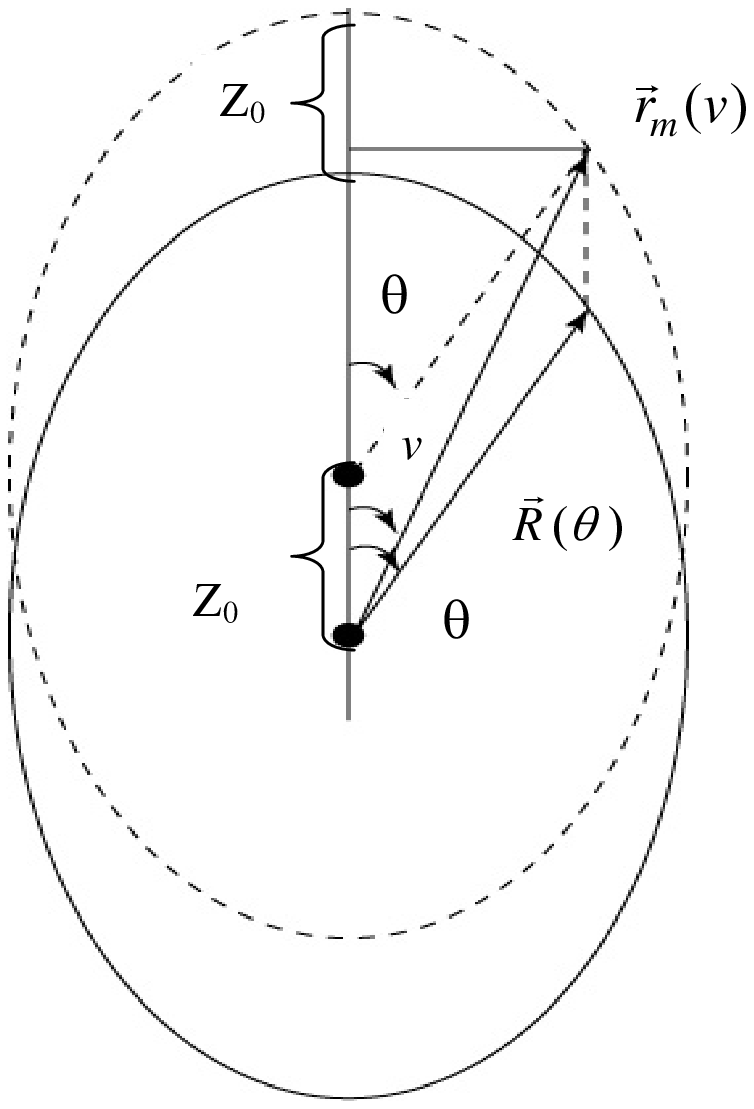}
\vskip-3mm\caption{  }
\end{figure}

As is seen from Eq.~(\ref{eq24}), $\cos\theta_{1}=0$ at $r=R_{\bot}.$
Therefore, Eq.~(\ref{eq31}) coincides with Eq.~(\ref{eq30}). At $r=R_{\Vert}$,
we have $\cos\theta_{1}=1$, and expression (\ref{eq31}) coincides with
(\ref{eq32}).

Passing from the ellipsoidal shape to the limiting case of the spherical
shape ($R_{\bot}=\,R_{\parallel}\equiv R$), i.e. at
$e_{p}\rightarrow0$, it is easy to see that $\Psi(n)\rightarrow0$
for all $n\neq0$. Concerning $\Psi(0)$,
Eqs.~(\ref{eq30})--(\ref{eq32}) show that, at this limiting transition,
\begin{equation}
\Psi(0)\rightarrow\Psi^{(s)}=V\,\rho_{i}\;\left\{  {%
\begin{array}
[c]{l l}%
{\left(  {3-\frac{r^{2}}{R^{2}}}\right)  }%
/{2R,}&{\mathrm{at}\quad r<R}\\
\frac{1}{r},& \mathrm{at}\quad r>R\\
\end{array}
}\right\},  \label{eq33}%
\end{equation}
where $V=\frac{4\pi}{3}R^{3}$ is the sphere volume.
Expression for $\Psi^{(s)}$ in the form (\ref{eq33}) was used in
\cite{2}, while considering nonlinear plasma oscillations in a
spherical metal nanoparticle.

\section{Electrostatic Force}

Provided that the field $\mathbf{E}_{L}(t)$ is oriented along the
axis $0Z,$ and the particle shape has the adopted symmetry, the field
$\langle F\rangle$ is also directed along the axis $0Z$, i.e.,
\[
\langle F\rangle =-eE_{L}(t)\,N_{e} +
\]
\begin{equation}
+n_{0}\,e\,\int{d^{3}r\,\mathbf{K}_{0}\,\boldsymbol{\nabla
}}\;\Phi_{i}^{(0)}(\mathbf
{r})\,\Delta_{0}\,\left(  {R(\theta)-\left\vert {\,\mathbf{r}-\mathbf{z}_{0}%
\,}\right\vert \,}\right),  \label{eq34}%
\end{equation}
where $\mathbf{K}_{0}$ is a unit vector directed along the axis
$0Z$. The integration range over $\mathbf{r}$ in Eq.~(\ref{eq34}) is
defined by the condition that the argument in the step-like function
$\Delta_{0}$ is larger than or equal to zero, i.e.,
\begin{equation}
R(\theta)-\left\vert {\,\mathbf{r}-\mathbf{z}_{0}}\right\vert \geq0.\label{eq35}%
\end{equation}
In the case where relation (\ref{eq35}) is the equality, we obtain a root
\begin{equation}
r\equiv r_{m}(\nu)=\left\{  {\,R^{2}(\theta)-z_{0}^{2}\;\sin^{2}\nu
\,}\right\}
^{1\mathord{\left/ {\vphantom {1 2}} \right. \kern-\nulldelimiterspace}2}%
+z_{0}\,\cos\nu .\label{eq36}%
\end{equation}

As is seen from Fig.~2, the vector $\mathbf{r}_{m}(\nu)$ corresponds
to that point on the surface of a shifted spheroid, which is
determined by the vector $\mathbf{R}(\theta)$ on the surface of
the initial spheroid. At the shift, the vector $\mathbf{R}(\theta)$
moves in parallel to itself from point 0 to point $0^{\prime}$. According to
Fig.~2, we can write
\begin{equation}
\mathbf{r}_{m}(\nu)\,\cos\nu=z_{0}+R(\theta)\,\cos\theta ,\label{eq37}%
\end{equation}
\begin{equation}
\mathbf{r}_{m}(\nu)\,\sin\nu=R(\theta)\,\sin\theta .\label{eq38}%
\end{equation}
Multiplying relations (\ref{eq37}) and (\ref{eq38}) by $\sin\nu$ and $\cos\nu
$, respectively, and subtracting the results, we obtain
\begin{equation}
z_{0}\,\sin\nu=R(\theta)\;\sin(\theta-\nu).\label{eq39}%
\end{equation}
This formula gives a relation between the angles $\theta$ and $\nu$ at a fixed
$z_{0}$. At $z_{0}\rightarrow0$, we obtain $\theta\rightarrow\nu$. Since the
ratio $z_{0}/R(\nu)$ is small, we can write
\begin{equation}
\theta=\nu+\Delta\,\nu \label{eq40}%
\end{equation}
so that $\Delta\nu$ can be determined from Eq.~(\ref{eq39}) by iterations:
\[
\Delta\nu_{1}=\frac{z_{0}}{R(\nu)}\sin\nu;\]
\[
\Delta\,\nu_{2}=\frac{z_{0}%
}{R(\nu)}\sin\nu\left( {1-\frac{{R}^{\prime}(\nu)}{R(\nu)}\Delta
\nu_{1}}\right);
\]%
\begin{equation}
\Delta\,\nu_{3}=\frac{z_{0}}{R(\nu)}\sin\nu\left(  {1-\frac
{{R}^{\prime}(\nu)}{R(\nu)}\Delta\nu_{2}}\right),\ldots~.\label{eq41}%
\end{equation}
To obtain the explicit dependence of $\mathbf{r}_{m}(\nu)$ on the
shift $z_{0}$, let us expand Eq.~(\ref{eq36}) in a series (the
function $R(\theta)$ should also be expanded with the use of relations
(\ref{eq40}) and (\ref{eq41}))
\[
r_{m}(\nu)=R(\nu)+z_{0}(1-e_{p}^{2})\frac{\cos\nu}{1-e_{p}^{2}\cos^{2}\nu
}-
\]
\[
-z_{0}^{2}\;\frac{1-e_{p}^{2}}{2R_{\perp}}\frac{\sin^{2}\nu}{(1-e_{p}
^{2}\cos^{2}\nu )^{3/2}}+
\]
\begin{equation}
+z_{0}^{3}\,\frac{e_{p}^{2}}{2R_{\bot}^{2}}\left( {\frac{1}{3}+\frac
{1-e_{p}^{2}}{(1-e_{p}^{2}\,\cos^{2}\nu)^{2}}}\right)  \,\cos\nu\,\sin^{4}%
\nu\,+\ldots~.
\label{eq42}%
\end{equation}
For the spherical shape ($e_{p}\rightarrow0$), Eq.~(\ref{eq36}) yields
\begin{equation}
\mathbf{r}_{m}(\nu)=R+z_{0}\cos\nu-\frac{z_{0}^{2}}{2R}\sin^{2}\nu-\frac
{z_{0}^{4}}{8R^{3}}\sin^{4}\nu+\ldots~.\label{eq43}%
\end{equation}
Comparing expressions (\ref{eq42}) and (\ref{eq43}), we see that, in the case
of an asymmetric particle ($e_{p}\neq0$), an extra cubic nonlinearity
absent in expression (\ref{eq43}) emerges.

To avoid a misunderstanding, we should emphasize that, in what follows,
the term \textquotedblleft asymmetric particle\textquotedblright, will mean a
particle, whose shape differs from the spherical one, rather than the absence
of any symmetry elements.

Note that formula (\ref{eq34}) with regard for Eqs.~(\ref{eq35}) and
(\ref{eq36}) can be written in the form
\[
\langle F\rangle =-eE_{L}(t)\,N_{e}+
\]
\begin{equation}
+2\pi\,n_{0}\,e\,\int\limits_{0}^{\pi}{d\nu\sin\nu}\int\limits_{0}^{r_{m}%
(\nu)}{dr\, r^{2}}\left[  {\mathbf{K}_{0}\,\boldsymbol{\nabla }\,\Phi_{i}^{(0)}%
(r)}\right] . \label{eq44}%
\end{equation}
Expressing $r_{m}(\nu)$ as
\begin{equation}
\mathbf{r}_{m}(\nu)=R(\nu)+\Delta r(\nu),\label{eq45}%
\end{equation}
and comparing it with Eq.~(\ref{eq42}), we see that only $\Delta r(\nu)$
depends on the shift $z_{0}$, and $\Delta r\ll R(\nu)$. It would seem that
the integral over $r$ in Eq.~(\ref{eq44}) can be expanded in a
power series in powers of $\Delta r_{m}(\nu)$ to obtain terms, both linear and nonlinear
in $z_{0}$. However, there is a subtle point. In order to obtain terms
nonlinear in $z_{0}$, we must differentiate the integrand in Eq.~(\ref{eq44}),
i.e. the function
\begin{equation}
w(r)=r^{2}\left[
{\mathbf{K}_{0}\boldsymbol{\nabla }\,\Phi_{i}^{(0)}}\right].\label{eq46}%
\end{equation}
However, as is seen, e.g., from the exact expression for the ion electrostatic
potential in the spherical case (\ref{eq33}), the function $\Psi^{(s)}(r)$ and
its first derivative are continuous at the point $r=R$, i.e. across the
particle surface. But already the second derivative of $\Psi^{(s)}(r)$ is
discontinuous at this point. Therefore, the mentioned integral cannot be
expanded in a Taylor series at the surface. At the same time, the ion
electrostatic potential is described by smooth functions to the left and to
the right from the surface. Therefore, we will do as follows. Let us divide
the integration path over $\nu$ in Eq.~(\ref{eq44}) into intervals, in which
$r_{m}(\nu)$ lies only to the left or only to the right from the nanoparticle
surface. Then, to the left and to the right from the surface, there exist
the eligible reasons for function (\ref{eq46}) to be expanded into a Taylor series.

Let us explain the essence of our approach using, as an example, the spherical
shape, for which the results are already known \cite{1,2}. Hence, the integral
entering Eq.~(\ref{eq44}) can be rewritten as follows:
\[
 \int\limits_{0}^{\pi}{dv\,\sin
v\int\limits_{0}^{r_{m}(v)}{dr\,r^{2}\left[
{\mathbf{K}_{0}\boldsymbol{\nabla }\Phi_{i}^{(0)}}\right]  }}=
\]
\begin{equation}
=\int\limits_{0}%
^{\pi/2}%
{dv\,\sin v\int\limits_{0}^{r_{m}(v)}{dr\,w(r)+\int\limits_{\pi
/2}^{\pi}{dv\,\sin v\int\limits_{0}^{r_{m}(v)}{dr\,w(r)}}}}.\label{48}%
\end{equation}
As is seen from Eq.~(\ref{eq43}), $r_{m}(\nu)\geq R$ for the first integral,
and $r_{m}(\nu)\leq R$ for the second one (these inequalities become somewhat
violated at $v\approx\pi/2$, but this does not make an appreciable
contribution to the integral). In the case of the spherical shape in accordance
with Eqs.~(\ref{eq33}) and (\ref{eq36}), we have
\begin{equation}
w(r)=-\rho_{i}\,V\,\cos\nu,\quad \mathrm{at}~~r\geq R,\label{eq47}%
\end{equation}
\begin{equation}
w(r)=-\rho_{i}\,V\,\cos\nu\left(  {\frac{r}{R}}\right)  ^{3},\quad
\mathrm{at}~~r\leq R.\label{eq48}%
\end{equation}
Now, let us substitute functions (\ref{eq47}) and (\ref{eq48}) into the first
and the second integral, respectively, on the right-hand side of expression
(\ref{eq46}). Then, let us take into consideration that $\mathbf{r}_{m}%
(\nu)=R+\Delta r(\nu)$, with $\Delta r(\nu)\ll R$, and make the expansion
\[
\int\limits_{0}^{r_{m}(\nu)}{dr\, r^{2}}\;\left[
{\mathbf{K}_{0}\,\mathbf {\nabla}\,\Phi_{i}^{(0)}(r)}\right] =
\]
\[
\!\! =\!\!\int\limits_{0}^{R+\Delta r(v)}
\!\!\!{dr\,w(r)=\!\int\limits_{0}^{R}{dr\,w(r)\!\!+w(r)\!\Delta
r}}\!+\!\frac{1}{2}\left(  {\frac{dw}{dr}}\right) _{R}\!(\Delta
r)^{2}+
\]%
\begin{equation}
+\frac{1}%
{6}\left(  {\frac{d^{2}w}{dr^{2}}}\right)  _{R}(\Delta r)^{3}+\frac{1}%
{24}\left(  {\frac{d^{3}w}{dr^{3}}}\right)  _{R}(\Delta
r)^{4}+\ldots~.
\label{eq49}%
\end{equation}
To avoid the misunderstanding, we note once again that the derivatives in
Eq.~(\ref{eq49}) are not calculated at the surface ($r=R$), but at a point
$r$, provided that $r$ approaches the surface from the left or from the right.
In particular, in the given specific case, the matter concerns the expansion
of function (\ref{eq48}), provided that $r\rightarrow R$ and $r\leq R$.

From expressions (\ref{eq42}) and (\ref{eq43}), we see that $\Delta
r$ is a power series in $z_{0}$, i.e.
\begin{equation}
\Delta r=\Delta r_{1}+\Delta r_{2}+\Delta r_{3}+\Delta r_{4},\label{eq50}%
\end{equation}
where $\Delta r_{i}\sim z_{0}^{i}$. In particular, in the case of spherical
symmetry, according to Eq.~(\ref{eq43}), we have
\[
\Delta r_{1}=z_{0}\cos v,\quad\Delta r_{2}=-\frac{z_{0}^{2}}{2R}\sin
^{2}v,
\]
\begin{equation}
\Delta r_{3}=0,\quad\Delta r_{4}=-\frac{z_{0}^{4}}{8R^{3}}\sin
^{4}v.\label{eq51}%
\end{equation}
Taking into account Eqs.~(\ref{eq49}) and (\ref{eq50}), all integrals in
expression (\ref{eq44}) can be calculated to the end, and we obtain (at
$z_{0}\geq0$)
\begin{equation}
\langle F\rangle=-e\,E_{L}
(t)\,N_{e}-m_{e}\,N_{e}\omega_{p}^{2}\left\{
{z_{0}-\frac{9}{16}\,\frac{z_{0}^{2}}{R}+\frac{z_{0}^{4}}{32R^{3}}}\right\}
,\label{eq52}%
\end{equation}
where $\omega_{p}^{2}=\frac{4\pi\,e^{2}n_{0}}{3\,m_{e}}$ is the square of
the plasma (dipole) frequency.

Thus, we repeated the result of works \cite{1,2}. Now, let us apply the same
approach to the case of a spheroidal nanoparticle.

\section{Nonlinear Dipole Plasma Oscillations of Asymmetric Metal
Nanoparticle}

In the case of a spheroidal nanoparticle with regard for
Eq.~(\ref{eq29}), we have
\[
\mathbf{K}_{0}\boldsymbol{\nabla }\Phi_{i}^{(0)}=\cos v\frac{\partial\Phi_{i}^{(0)}%
}{\partial r}-\frac{\sin v}{r}\,\frac{\partial\Phi_{i}^{(0)}}{\partial
v}=
\]
\[
=\cos v\biggl\{  \frac{\partial\Psi(0)}{\partial
r}+\frac{\partial\Psi (2)}{\partial r}P_{2}(\cos v)+
\]
\begin{equation}
+\frac{2}{r}\Psi(2)\left[  1-P_{2}(\cos
v)\right]  +\ldots\biggr\}.  \label{eq53}%
\end{equation}
Below, we take into account explicitly only $\Psi(0)$,
although the required calculations were also carried out making
allowance for the contribution of the function $\Psi(2)$, the expression
for which is presented in Appendix. Our estimates showed that the
account of the contribution made by $\Psi(2)$ does correct, to some
extent, the coefficients at the powers of $z_{0}$, but does not
change our main conclusions.

Hence, in accordance with Eq.~(\ref{eq42}), we have
\begin{equation}
\label{eq54} r_m (0)= R_{\parallel } +z_0 >R_{\perp} ;\quad r_m (\pi
)=R_{\parallel} -z_0 < R_{\parallel }.
\end{equation}
Taking these inequalities into account, let us split the integral
over the angle $\nu$ again as was done in Eq.~(\ref{48}). We now
substitute function (\ref{eq32}) in the first integral on the
right-hand side of Eq.~(\ref{48}); here, in accordance with
Eq.~\ref{eq54}, $r_{m}(0)>R_{\parallel}.$ Then, according to
Eqs.~(\ref{eq46}) and (\ref{eq29}), we have
\begin{equation}
\label{eq55} w(r)=\cos v\, \,r^2\frac{\partial \Psi (0)}{\partial
r}=-\frac{4\pi }{3}\,\rho _{i\,} R_\perp ^2 \;R_{\parallel} \cos v.
\end{equation}
In the second integral in Eq.~(\ref{48}), for which
$r_{m}(\pi)<R_{\parallel}$, we also suppose that
$r_{m}(v)>R_{\perp}$ and use function (\ref{eq31}). In this case, we
obtain
\begin{equation}
w(r)=\frac{4\pi}{3}\,\rho_{i\,}\left\{  {\frac{1}{e_{p}}(r^{2}-R_{\bot}%
^{2})^{3/2}%
-r^{3}}\right\}  \cos v.\label{eq56}%
\end{equation}
Note that the assumption $r_{m}(v)>R_{\perp}$ also means that
\begin{equation}
r(\pi)= R_{\parallel}-z_{0}>R_{\perp}~~\text{or} ~~z_{0}<
R_{\parallel}-R_{\perp
} .\label{59}%
\end{equation}
Without assumption (\ref{59}), the expressions for the ion
electrostatic potential, as well as the integration limits
(\ref{eq43}), are transformed into the corresponding result for a
spherical particle at $e_{p}\rightarrow0$. Condition (\ref{59})
makes this passage to the limit impossible, because, if
$e_{p}\rightarrow0$, inequality (\ref{59}) becomes invalid at any
small, but finite value of $z_{0}$. In this case, for the passage to
the limit $e_{p}\rightarrow0$ to be eligible, one should engage
function
(\ref{eq30}) rather than function (\ref{eq31}). Hence, the substitution of Eqs.~(\ref{eq55}%
) and (\ref{eq56}) in Eq.~(\ref{48}) gives
\[
\int\limits_{0}^{\pi}{dv\sin v}\int\limits_{0}^{r_{m}(v)}{dr \left[
{\mathbf{K}_{0}\boldsymbol{\nabla}\Phi_{i}^{(0)}}\right] } =
\]
\[
={{-\frac{4\pi}{3}R_{\perp}%
^{2}\,R_{\parallel}\rho_{i}\,\int\limits_{0}^{\pi/2}%
{dv\,r_{m}(v)\,\sin v\,\cos v\;+}}}
\]
\begin{equation}%
+\frac{4\pi}{3}\rho_{i}\!\!\int\limits_{\pi/2}^{\pi }\!\!{dv\sin
v\cos v}\!\!\!
{\int\limits_{0}^{r_{m}(v)}\!\!\!{dr\left\{ \! {\frac{1}{e_{p}%
}(r^{2}\!-R_{\bot}^{2}%
)^{3/2}%
-r^{3}}\!\right\}  }}.
\label{eq57}%
\end{equation}
To integrate the second integral over $r$ in Eq.~(\ref{eq57}), we
take advantage, similarly to Eq.~(\ref{eq49}), of the smallness of
quantity $\Delta r$ and expand this integral in a series in $\Delta
r$. However, there exists a certain difference between cases
(\ref{eq49}) and (\ref{eq57}). In case (\ref{eq49}),
$r_{m}(v)=R+\Delta r(v),$ and in case (\ref{eq57}),
$r_{m}(v)=R(v)+\Delta r(v)$ in accordance with Eq.~(\ref{eq42}). To
avoid
excess complications, we expand the integral $\int_{0}^{r_{m}%
(v)}{dr\,\left\{\ldots\right\}  }$ into a series in $\Delta r$ at
the point $R_{\parallel},$ rather than at $R(v)$. A reason for this
approximation is that, first, the electrostatic potential is mainly
governed by the distribution of charges near the ellipsoid vertex
(pole), i.e. by the range of angles, in which $R(v)\approx
R_{\parallel}$, and, second, the function $\Psi(0)$ and its
derivative, as is seen from Eqs.~(\ref{eq31}) and (\ref{eq32}), are
continuous at the point $r=R_{\parallel}$, similarly to what takes
place in the spherical case, for which the exact solution is known.

From Eq.~(\ref{eq57}), confining the expansion to terms cubic in $\Delta r$,
we obtain
\[
\int\limits_{0}^{\pi}{dv\,\sin
v}\int\limits_{0}^{r_{m}(v)}{dr\,r^{2}\left[
{\mathbf{K}_{0}\boldsymbol{\nabla}\Phi_{i}^{(0)}}\right]} =
\]
\[
={{-V\,\rho_{i}\int \limits_{0}^{\pi
/2}%
{dv\,\Delta r(v)\sin v\cos v}}}-
\]%
\begin{equation}
-V\rho_{i}\int\limits_{\pi /2}^{\pi
}{dv\,\sin v\,\cos v\,\left\{  {\Delta r(v)-\frac{(\Delta r(v))^{3}}%
{2e_{p}^{2}\,R_{\parallel}^{2}}}\right\}  }, \label{eq58}%
\end{equation}
where $V=\frac{4\pi}{3}R_{\perp}^{2}R_{\parallel}$ is the volume of
spheroid.

Note that, owing to inequality (\ref{59}), the following inequality, as can be
easily verified, is also valid:
\begin{equation}
\label{eq59} e_p\, R_{\parallel} >z_0.
\end{equation}
If inequality (\ref{eq59}) is obeyed, the term cubic in $\Delta r$
is much smaller than the linear one, as it must be when expanding in
a small parameter. Since the expression for $\Delta r(v)$ itself is
a series expansion in $z_{0}$, we confine the consideration below to
the terms, the order of which is not higher than $z_{0}^{3}$, i.e.
we make the substitution
\[
\Delta r(v)\approx\Delta r_{1}(v)+\Delta r_{2}(v)+\Delta r_{3}(v),
\]
\begin{equation}
\left( {\Delta r(v)}\right)  ^{3}\approx\left(  {\Delta
r_{1}(v)}\right)
^{3}\label{eq60}%
\end{equation}
into Eq.~(\ref{eq58}). The form of $\Delta r_{i}(v)$-terms for the spheroidal
shape is clear from expression (\ref{eq42}).

Calculating the corresponding integrals in Eq.~(\ref{eq58}) and substituting
the obtained expression into Eq.~(\ref{eq44}), we obtain
\begin{equation}
\left\langle F\right\rangle =-e\,E_{L}(t)\,N_{e}-m_{e}N_{e}\,\omega_{\parallel}%
^{2}\,z_{0}-m_{e}\,N_{e}\,\omega_{pL}^{2}\,\frac{\delta(e_{p})}{R_{\parallel}^{2}%
}z_{0}^{3}, \label{eq61}%
\end{equation}
where
\begin{equation}
\omega_{\parallel}^{2}=L_{\parallel}\omega_{pL}^{2}\equiv\frac{1-e_{p}^{2}%
}{2e_{p}^{3}}\left\{  {\ln\left(  {\frac{1+e_{p}}{1-e_{p}}}\right)  -2e_{p}%
}\right\} \omega_{pL}^{2}, \label{eq62}%
\end{equation}
$\omega_{pL}=\sqrt{{\frac{4\pi n_{0}e^{2}}{m_{e}}}}$ is the plasma
frequency, and $L_{\parallel}$ is the depolarization factor along
the symmetry axis in the case of prolate spheroid $\left(
{R_{\parallel}>R_{\perp}}\right).$ In addition,
we introduced a dimensionless parameter $\delta(e_{p})$ in Eq.~(\ref{eq61}), which
depends only on the eccentricity $e_{p}$ and looks like
\[
\delta(e_{p}) =\frac{4}{315}\frac{e_{p}^{2}}{1-e_{p}^{2}}-\frac
{13}{12e_{p}^{2}}+\frac{5}{2e_{p}^{4}}-
\]
\[
-\frac{5-6e_{p}^{2}+e_{p}^{4}}%
{8e_{p}^{5}}\ln\left(  \frac{1+e_{p}^{{}}}{1-e_{p}}\right) -
\]
\begin{equation}
-\frac{1-e_{p}^{2}}{16e_{p}^{4}}\left\{  \frac{5}{2}-\frac{3}{2e_{p}^{2}%
}+\frac{3(1-e_{p}^{2})^{2}}{4e_{p}^{3}}\,\ln\left(  \frac{1+e_{p}}{1-e_{p}%
}\right)  \right\} . \label{eq63}%
\end{equation}
Comparing the expressions obtained for the electrostatic force in
the cases of spherical (formula (\ref{eq52})) and ellipsoidal
(formula (\ref{eq61})) nanoparticles, we see that, for the
asymmetric particle, the quadratic nonlinearity changes to the cubic
one. We recall once more that expression (\ref{eq63}) was obtained
in the assumption $0<e_{p}<1$, and, therefore, the passage to the
limit $e_{p}\;\rightarrow0$ or $e_{p}\;\rightarrow1$ cannot be
justified. To get some understanding concerning the magnitude of
parameter $\delta(e_{p})$, we give the following values:
\[
\delta\left(\frac{1}{2}\right)=-\frac{1}{7},\quad\delta\left(\frac{1}{5}\right)=-1.
\]
In our case, when oscillations occur along the symmetry axis (${\mathbf{u}%
=\mathbf{z}_{0}}$), the equation of motion (\ref{eq5}) with regard
for expression (\ref{eq61}) reads
\begin{equation}
\ddot{{z}}_{0}+\omega_{\parallel}^{2}\;z_{0}+\omega_{pL}^{2}\;\frac{\delta(e_{p}%
)}{R_{\parallel}^{2}}\;z_{0}^{3}=-\frac{e\;E(t)}{m_{e}}.\label{eq64}%
\end{equation}
The frequency $\omega_{\Vert}$ corresponds to the frequency of a dipole plasmon,
when the dipole oscillates along the symmetry axis of the spheroid. If we put
\begin{equation}
E_{L}(t)=E_{0}\;\cos\omega t\label{eq65}%
\end{equation}
and assume that the nonlinearity is weak, Eq.~(\ref{eq64}) can be solved using
the iteration method:
\[
z_{0}\approx\frac{e\,E_{0}}{m}\;\frac{\cos\omega\,t}{\omega_{\parallel}^{2}%
-\omega^{2}}+\frac{\omega_{pL}^{2}\,\delta(e_{p})}{4R_{\parallel}^{2}}
\;\frac{\left(  {{e\,E_{0}}m_e
}\right)^{3}}{(\omega_{\parallel}^{2}-\omega^{2})^{3}}\times
\]
\begin{equation}
\times\left\{ {\frac
{3\,\cos\omega\,t}{\omega_{\parallel}^{1}-\omega^{2}}+\frac{\cos3\omega\,t}%
{\omega_{\parallel}^{2}-(3\omega)^{2}}}\right\}.  \label{eq66}%
\end{equation}
In Eq.~(\ref{eq64}), the dissipation was not taken into account.
Therefore, solution (\ref{eq66}) has a singularity at
$\omega\rightarrow\omega_{\Vert}$. The insertion of a standard term
$2\gamma  {z}_0 $, which takes the oscillation attenuation into
account, into the left-hand side of Eq.~(\ref{eq64}) gives rise to a
disappearance of the singularity from the solution. In particular,
in the linear approximation, instead of the solution
$\frac{eE_{0}}{m}\frac{\cos\omega\,t}{\omega_{\Vert}^{2}-\omega^{2}}$,
we have $
\frac{eE_{0}}{m}\,\frac{\cos(\omega\,t-\varphi_{0})}{\left\{
{(\omega _{\parallel}^{2}-\omega^{2})^{2}+4\gamma^{2}}\right\}
^{1/2}},%
$
where $\varphi_{0}$ is the phase. Similar modifications must also be made in
the nonlinear terms. Note that Eq.~(\ref{eq64}) with $E_{L}(t)$ in the form
(\ref{eq65}) corresponds to the so-called Duffing equation. The analysis of
its solutions can be found, e.g., in work \cite{5}.

Now, having the explicit expression for the displacement of the
center of masses of  electrons, $z_{0}$, in terms of the laser field $E_{0}$ (see
Eq.~(\ref{eq66})), we can write down the formula for the dipole moment of a spheroidal metal
nanoparticle. If the laser field is polarized along the symmetry axis of
the spheroid, the dipole moment has the same orientation and equals
\begin{equation}
d=Ve\,n_{0}\,z_{0}=d_{1}+d_{2}.\label{eq67}%
\end{equation}
Here, $d_{1}$ is the linear dipole component, which, in accordance with
Eq.~(\ref{eq66}), equals
\begin{equation}
d_{1}=\;\frac{V}{4\pi}\;\frac{\omega_{pL}^{2}}{\omega_{\parallel}^{2}-\omega^{2}%
}\,E_{0}\,\cos\omega\,t. \label{eq68}%
\end{equation}
Similarly, the cubic component of the dipole, $d_{3}$, can be written down in
the form
\[
d_{3}=\frac{V\;\delta(e_{p})}{16\pi\,R_{\parallel}^{2}}\,\frac
{({e m})^{2}%
\omega_{pL}^{3}}{(\omega_{\parallel}^{2}-\omega^{2})^{3}}\,E_{0}^{3}\times
\]
\[
 \times\left\{
{\frac{3\cos\omega\,t}{\omega_{|\,|}^{2}-\omega^{2}}+\frac{\cos3\omega
t}{\omega_{\parallel}^{2}-(3\omega)^{2}}}\right\}  .
\]
At last, we would like to make the following remark. If the
dimensionless displacement $\alpha=z_{0}/R_{\Vert}$ is introduced,
Eq.~(\ref{eq64}) looks like
\begin{equation}
\ddot{\alpha}+\omega_{\parallel}^{2}\;\alpha+\omega_{pL}^{2}\,\delta(e_{p}%
)\,\alpha^{3}=\frac{eE(t)}{mR_{\parallel}}. \label{eq69}%
\end{equation}
We see that the magnitude of nonlinearity is determined by the dimensionless
parameter $\delta(e_{p})$, the analytic form of which, as a function of
the eccentricity $e_{p},$ is given by formula (\ref{eq63}). In addition, the specific
$\delta(e_{p})$-values at $e_{p}=1/2$ and $1/5$ were quoted above. This allows
us to estimate the nonlinearity.

It is also worth noting that the nonlinearity considered above is induced by
the electric component of the laser wave field. Under certain conditions (the
particle size, the field frequency), the nonlinearity can be induced by the
magnetic component. In particular, the effect of second harmonic generation in
spherical metal particles under the influence of the magnetic component of the
laser wave field was considered in work \cite{6}.

\section{Conclusions}

It has been shown that the laser field oriented along the symmetry axis of a
spheroidal metal nanoparticle generates a cubic nonlinearity, which is
absent in the case of a spherical particle. Instead, the quadratic
nonlinearity inherent to the case of spherical symmetry disappears. An approximate
analytic expression for the dipole moment of a spheroidal metal nanoparticle
has been derived to within terms cubic in the field.

\subsubsection*{APPENDIX}%
{\footnotesize
\[
\Psi(2)=2\pi\rho_{i}r^{2}\,\left\{  {\frac{1}{3}-\frac{e_{p}%
^{2}-1}{e_{p}^{2}}+\frac{e_{p}^{2}-1}{2e_{p}^{3}}\ln\left(  {\frac{1+e_{p}%
}{1-e_{p}}}\right)  }\right\},
\]
\[
 \mathrm{at}\quad r\leq
R_{\perp}.
\]%
\[
\Psi(2)=2\pi\rho_{i}r^{2}\left\{\!\!  {%
\begin{array}
[c]{l}%
\frac{1}{3}(1-\cos^{3}\theta_{1})+\frac{1}{5}(1-\cos^{2}\theta_{1})\cos
\theta_{1}-\\[2mm]
-\frac{e_{p}^{2}-1}{e_{p}^{2}}(1-\cos\theta_{1})-\\[2mm]
-\frac{e_{p}^{2}-1}{2e_{p}^{3}}\ln\left[  {\frac{1+e_{p}\cos\theta_{1}%
}{1-e_{p}\cos\theta_{1}}\;\frac{1-e_{p}}{1+e_{p}}}\right] +\\[2mm]
+\frac{1}%
{15}\left[ \left(\!  {\frac{R_{\perp}}{r}}\!\right)^{2}\left(
{3\cos^{2}\theta_{1} \!-\!1}\right)  \!-\!2\left(\!
{\frac{R_{\perp}}{r}}\!\right)^{4} \right]
\cos\theta_{1}\\[2mm] \text{при}\quad R_{\perp}\leq r\leq R_{\parallel}\\
\end{array}
}\!\!\!\right\},
\]%
\[
\Psi(\ref{eq2})=\frac{4\pi}{15}\rho\,_{i}R_{\perp}^{2}\,e_{p}^{2}\,\left(
{\frac{R_{\parallel}}{r}}\right)  ^{3},~~~\mathrm{at}\;\;r\geq R_{\parallel}.%
\]

}

\rezume{%
НЕЛІНІЙНІ ПЛАЗМОВІ ДИПОЛЬНІ КОЛИВАННЯ\\ У СФЕРОЇДАЛЬНИХ МЕТАЛЕВИХ
НАНОЧАСТИНКАХ}{П.М. Томчук, Д.В. Бутенко} {У роботі розвинуто теорію
нелінійних дипольних плазмових коливань у металевій наночастинці
сфероїдальної форми, які генеруються полем лазерної хвилі. Для
випадку лазерного поля, орієнтованого вздовж осі обертання сфероїда,
отримано наближені аналітичні вирази для дипольного моменту
наночастинки (з точністю до кубічної складової).}

\end{document}